\documentclass[10pt,a4paper]{article}
\usepackage{amsmath}
\usepackage{graphicx}
\usepackage[pagestyles]{titlesec}
\titleformat{\abstract}{\filcenter\normalfont\normalsize\bfseries\MakeUppercase}
{\theabstract.}{1em}{\normalsize}
\titleformat{\section}{\normalfont\normalsize\bfseries\MakeUppercase}
{\thesection.}{1em}{\normalsize}
\titleformat{\subsection}{\normalfont\normalsize\bfseries\itshape}{\thesubsection.}{1em}{\normalsize}
\titleformat{\subsubsection}{\normalfont\normalsize\itshape}{\thesubsubsection.}{1em}{\normalsize}
\begin{document}
\begin{titlepage}
\pagestyle{empty}
\begin{center}
{\large{\textbf{Low Cost PC Based Real Time 
Data Logging System Using PCs Parallel Port For Slowly Varying Signals}}}
\bigskip\bigskip \\

N. Monoranjan Singh*, K. C. Sarma\\

\bigskip

\textit{Department of Instrumentation \& USIC, Gauhati University\\ Guwahati – 781014, Assam (India)\\ *Email:- mranjansn@yahoo.com}\\

\vspace{1.5 cm}

\end{center}
\abstract
\bigskip
A low cost PC based real time data logging system can be used in the laboratories for the measurement, monitoring and storage of the data for slowly varying signals in science and engineering stream. This can be designed and interfaced to the PCs Parallel Port, which is common to all desktop computers or Personal Computers (PCs). By the use of this data logging system one can monitor, measure and store data for slowly varying signals, which is hard to visualise the signal waveforms by ordinary CRO (Cathode Ray Oscilloscope) and DSO (Digital Storage Oscilloscope). The data so stored can be used for further study and analysis. It can be used for a wide range of applications to monitor and store data of temperature, humidity, light intensity, ECG signals etc. with proper signal conditioning circuitry.\\ 

\end{titlepage}

\newpage
\setcounter{page}{1}
\section{INTRODUCTION}
It is very much essential in case of some industrial as well as experimental setup to monitor and control temperature, light intensity, humidity or biomedical signals continuously, which are of slowly varying signals. The efficient solution for such a problem is to develop a data acquisition system. The early development of data logger was done through manual measurements from analog instruments such as thermometers and manometers. Unfortunately this type of data logger can’t fulfill the current requirements in terms of time and accuracy. From 1990 a further development in data logging took place as people begin to create PC based data logging systems. These systems combine the acquisition and storage capabilities with archiving, reporting and display capabilities [\ref{1}]-[\ref{3}]. The waveform of such slowly varying signals cannot be displayed by ordinary Oscilloscope and DSO. Using the DAQ system, such a waveform can be displayed into the monitor on real time mode and the data can also be stored into the hard disc of the PC for further analysis. By designing an interfacing circuit using ADC 0808, the measured temperature in the analog form is first converted into digital (TTL level voltages) and feed to the PCs Parallel Port. A ``C - program'' is written to monitor and control the data acquisition system and also allows the data to save into a file into .csv format (Comma Separated Value). The program also calculates the relative humidity and dew points. By taking the difference in the readings of the temperature of the room from dry and wet sensors, relative humidity of the room can also be recorded automatically. The file so saved can be directly open with Microsoft Excel for further study and analysis.\\

Monitoring and controlling physical parameters like temperature, pressure, humidity light etc. by embedded system using microcontrollers are very much effective in industrial and research oriented requirements [\ref{4}]. In the present work, the design of low cost PC based data acquisition in hardware and software are made so as to make compatible in both new and legacy hardware(s) for desktop and laptops.\\

In general, the measurement of signal(s) requires using a transducer, a device that can convert one form of energy or signal into another form. Such a signal is generally analog, and is to be digitized using ADC. Then the digital data can be feed to the PC through the parallel ports. Then the data so sent can be received by the PC and can display graphically and stored into a file into the hard disk for future analysis.

\section{DESIGN DETAILS}
\subsection{Hardware Design}
As per the sampling theorem, the signal should be sampled at a frequency greater than twice the maximum frequency component present in the signal. If the signal is under sampled (i.e. sampled at a frequency less than twice the maximum frequency component in the signal) aliasing will occur. Aliasing refers to reflection of high frequency components into low frequencies in the frequency spectrum. Aliasing results in error in frequency spectrum computation. To prevent aliasing, an anti-aliasing filter (a low-pass filter) is used at the input stage. The input signal is passed through the filter. As the temperature of the room does not change abruptly, we make a sample rate of two samples per second. Sample rate can be changed by slide modification in the program according to our choice.\\

ADC 0808 is used to convert the analog signals to digital data. ADC0808 is an 8-bit successive approximation type ADC and operates with +5 V supply voltage. It has $100\mu s$ conversion time at 640 kHz and operates within 10 kHz to 1280 kHz clock frequency [\ref{5}]. Successive approximation type ADCs require the analog input signal to be held constant.\\

A clock circuit is made using an IC 7414 and one ceramic capacitor and one resistor. The expression of the frequency generated is given in Fig. 1
\begin{center}
\begin{equation}
f=\dfrac{1}{1.1RC}
\end{equation}
\end{center}
\begin{center}
\begin{figure}[!h]
\centering
\includegraphics[width=4.5cm]{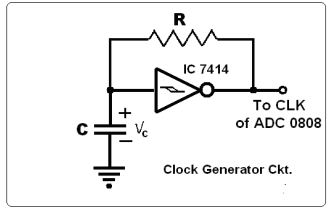}
\caption{Circuit diagram for the clock generator}
\end{figure}
\end{center}
\begin{figure}[!h]
\centering
\includegraphics[width=12cm]{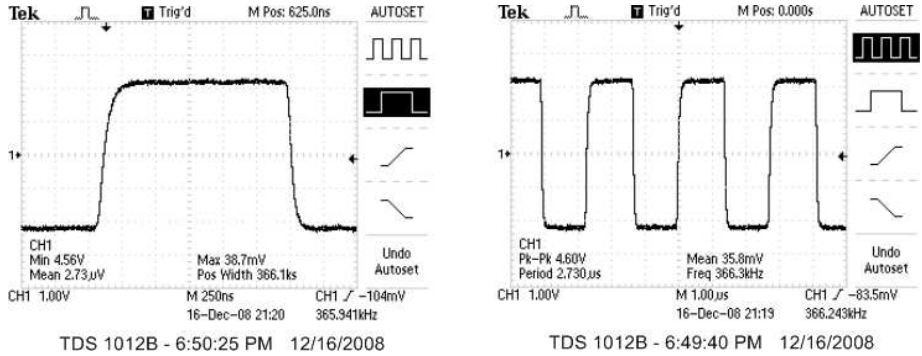}
\centering \caption{The waveforms generated by the clock circuit}
\end{figure}
The shape of the generated clock signal waveforms are shown above in Fig. 2. The waveform is captured using Tektronix DSO( Model No. TDS 1012B.)\\

The circuit diagram for the data acquisition is shown in fig 3. It consists of signal conditioning circuit, over voltage protection circuit, ADC circuitry, voltage level converter (if necessary) etc.\\
\begin{figure}[!h]
\centering \includegraphics[width=12cm]{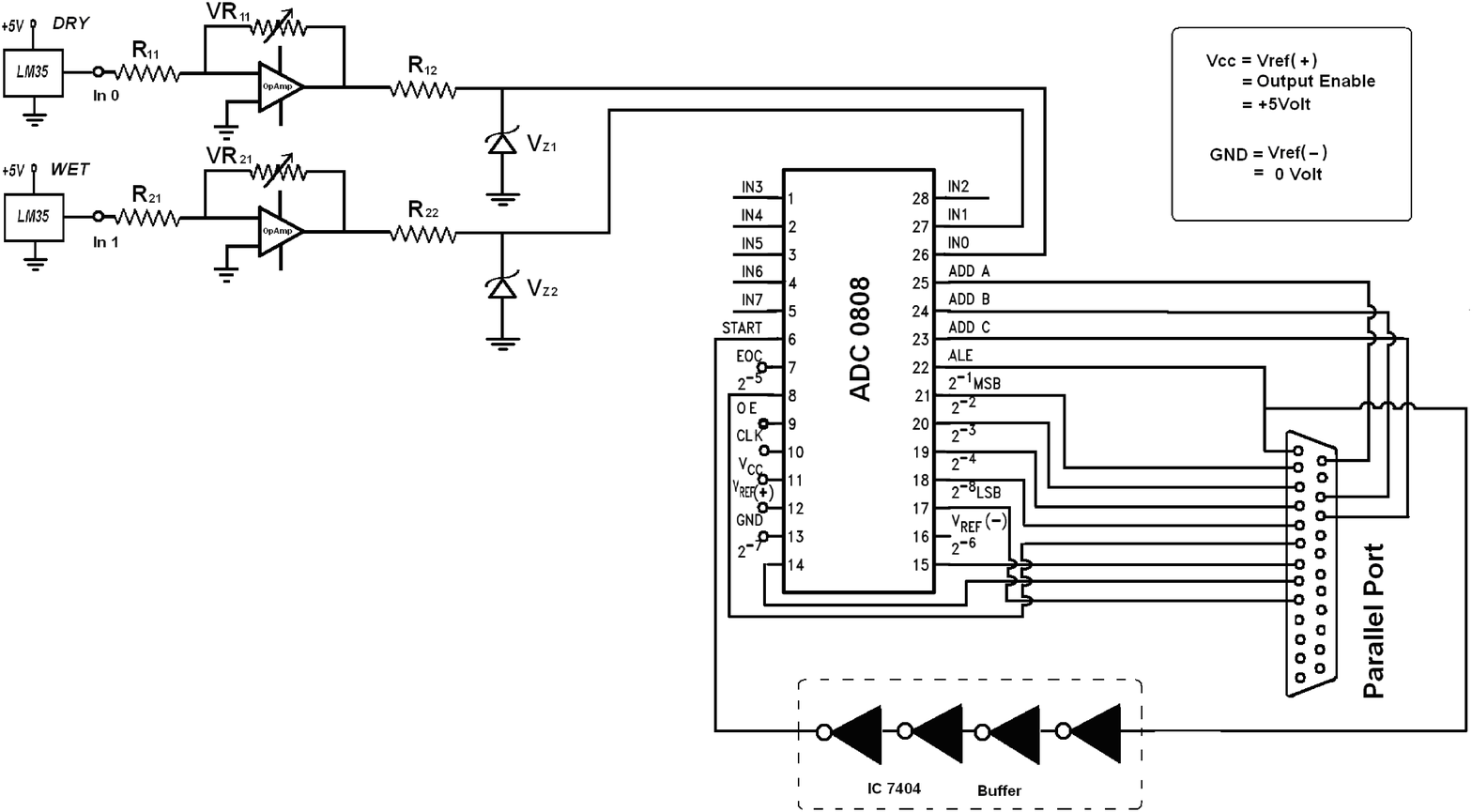}
\centering \caption{Circuit diagram for the data acquisition system}
\end{figure}
\subsection{Experimental Setup}
The circuit diagram of the experimental setup is shown in
Fig. 3. It consists of temperature sensor IC LM35, and an amplifier for setting proper adjustment and calibration, shunt zener diodes $V_{Z1}$ and $V_{Z2}$ for over range protection, an ADC 0808 and D25 parallel port connector.\\

The detector unit is the temperature sensor IC LM35, which is pre calibrated in $mV/^{\circ}C$ [\ref{6}]. The amplifier unit consists of an OpAmp of low offset and high gain. It amplifies the analog voltage obtained from the detector unit up to a certain desire voltage level (i.e. 5 volt max as the max input voltage of the ADC is 5 volt). If the sensor is kept beyond the specified range, the output at the amplifier will exceed the 5 volt limit and may generate error. This may lead to damage the ADC also. For safety purposes a zener diode is connected across the output of the amplifier. It limits the max output voltage to 5 volt and thus protects the ADC.\\

The input analog signal so obtained is converted into 8-bit digital signal by the ADC 0808 in 256 steps. Thus the resolution of the ADC is 0.0196 V ($\approx$ 19.6 mV). The digital data so obtained is fed into the PC through the parallel port. The operation of the ADC and Graphics Display with Data logging is controlled by using a programme written in ``C''. The display of the temperature in Graphical form is in real time mode and data is also saved in a file simultaneously for further use and analysis.\\
\subsection{Software Development}
Flowchart for the application program for data acquisition system is given in fig 4.
\begin{figure}[!h]
\centering \includegraphics[height=18cm]{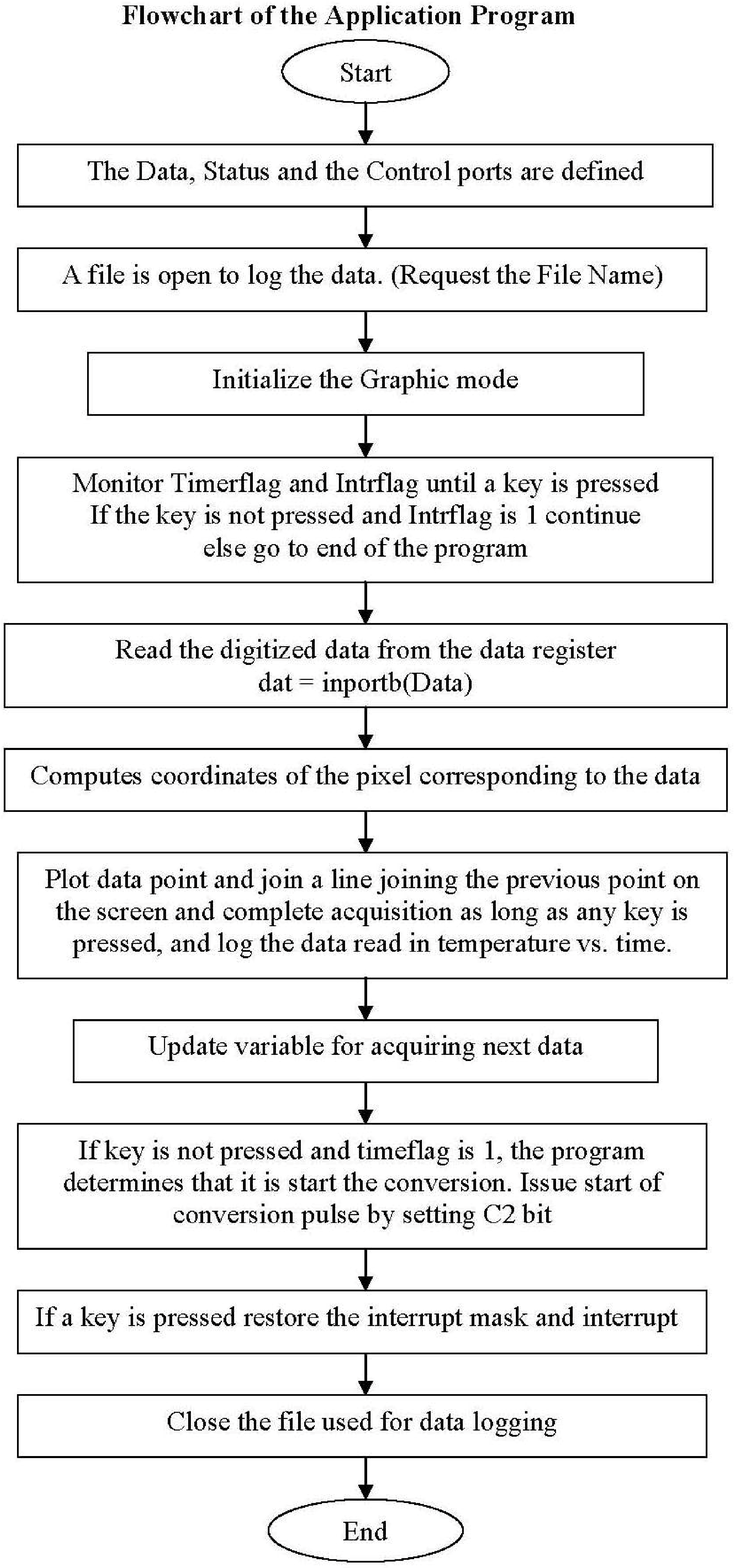}
\caption{Flow chart for the application  program}
\end{figure}
\subsubsection{Characteristics Graph of the LM35}
The characteristics graph of the LM35 (Fig 5) for calibration is made from the experimental observation. It is almost linear in characteristics. Basically LM35 is pre calibrated in degree celsius to be used for temperature sensor IC.
\begin{figure}[!h]
\centering \includegraphics[width=8cm]{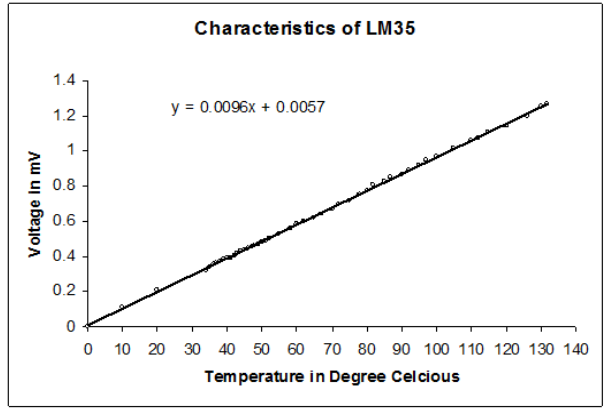}
\caption{Graph for the characteristics of the LM35}
\end{figure}

One can observe the temperature continuously from the onscreen graph of the room temperatures(in the 2$^{nd}$ minute) for the dry and wet sensors. The temperature and humidity are also calculated automatically in real time mode.\\

The recorded data of the average room temperature, relative humidity, and dew point is:\\
\begin{table}[!h]
\centering \begin{tabular}{|l|c|}
\hline
Dry Temp & 19.92858\\
\hline
Wet Temp & 18.02167\\
\hline
Rel. Humidity & 85.183416\\
\hline
Dew Point & 17.360743\\
\hline
\end{tabular}
\end{table}
\section{RESULT}
\subsection{Range and Resolution of the DAS}
The range of this data logger is from $0^{\circ} C$ to $50^{\circ} C$. This data acquisition system has a sample rate of two samples per second and a resolution temperature of $0.196 ^{\circ} C$ ( $\approx 0.2 ^{\circ} C$).
\subsection{Accuracy of the DAS}
Basically the LM35 series are precision integrated-circuit temperature sensors, whose output voltage is linearly proportional to the temperature in Celsius (Centigrade). Total unadjusted error for ADC 0808 is $\pm{\frac{1}{2}}$ LSB and $\pm 1$ LSB for ADC 0809 [\ref{5}]. And due to the presence of noise, poor shielding and some other factors accuracy is somewhat reduced. The data so obtained is compared with the reading from dry and wet temperatures from mercury thermometers and it gives quite satisfactory results.
\section{CONCLUSION}
The parallel port of the computer is usually used for printer and connecting modems because of its fast data transfer rate. It can be used for accepting digital data using an analog to digital converter or reading the data from the computer using a digital to analog converter [\ref{7}]. The data logger so designed can be used for real time measurements and monitoring for temperature, humidity etc. and storing data accurately for a long duration which is inconvenient for a manual observation. Also, by using appropriate sensor and transducers with relevant circuitry, it can be used for biomedical applications.
\bibliographystyle{plain}

\begin{thebibliography}{1}

\bibitem{}\label{1}  N. Monoranjan Singh, K. C. Sarma, ``Design of low cost PC based data acquisition system for real time temperature monitoring and data logging'', Journal of Gauhati University Research Scholars’ Association, Vol III, pp50-56, 2007-08
\bibitem{}\label{2} N. Monoranjan Singh, K. C. Sarma, ``Design of low cost 8-channel PC based data acquisition system using PCs parallel port'', presented in 96th Science Congress, pp13-14, Jan 3-7, 2009 Shillong, India.
\bibitem{}\label{3} National Instruments - A review of PC based data logging and recording techniques available at www.ni.com/datalogers.
\bibitem{}\label{4} A. Goswami, T. Bezboruah and K. C. Sarma, ``Design of an Embedded Systems for monitoring and controlling temperature and light'', Int. J. of Electronic Engineering Research, Vol 1 No. 1  pp27-36, .
\bibitem{}\label{5} National semiconductor corporation – ADC 0809 datasheet, oct 2002 updates.
\bibitem{}\label{6} National semiconductor corporation – LM35 datasheet, Nov 2000 updates.
\bibitem{}\label{7} Jan Axelson, ``Parallel Port Complete: Programming, Interfacing \& using the PC’s Parallel Printer Port'', Penram International Publishing (India) Pvt. Ltd

\end{thebibliography}

\end{document}